# Implementation and verification of different ECC mitigation designs for BRAMs in flash-based FPGAs[*]

YANG Zhen-Lei(杨振雷) [1,2]   WANG Xiao-Hui (王晓辉) [1]   ZHANG Zhan-Gang (张战刚)[3]
LIU Jie(刘杰) [1]   SU Hong(苏弘) [1;1)]

1 Institute of Modern Physics, Chinese Academy of Sciences, Lanzhou 730000, China

2 University of Chinese Academy of Sciences, Beijing 100049, China

3. Science and Technology on Reliability Physics and Application of Electronic Component Laboratory, China Electronic Product Reliability and Environmental Testing Research Institute, Guangzhou 510610, China

**Abstact:** Embedded RAM blocks (BRAMs) in field programmable gate arrays (FPGAs) are susceptible to single event effects (SEEs) induced by environmental factors such as cosmic rays, heavy ions, alpha particles and so on. As technology scales, the issue will be more serious. In order to tackle this issue, two different error correcting codes (ECCs), the shortened Hamming codes and shortened BCH codes, are investigated in this paper. The concrete design methods of the codes are presented. Also, the codes are both implemented in flash-based FPGAs. Finally, the synthesis report and simulation results are presented in the paper. Moreover, the heavy-ion experiments are performed, the experimental results indicate that the error cross-section using the shortened Hamming codes can be reduced by two orders of magnitude compared with the device without mitigation, and no errors are discovered in the experiments for the device using the shortened BCH codes.

## 1   Introduction

With the increased popularization of programmable logic, field programmable gate arrays (FPGAs) are gaining interest in space applications due to their re-programmability feature [1-3]. However, the sensitivity of FPGAs to single event effects (SEEs) is still a major concern in harsh radiation environment [4]. With technology evolving, supply voltage and node capacitance present a decline tendency [5]. As a result, the critical charge ($Q_{crit}$) is scaling down, which makes FPGAs become more susceptible to SEEs [6, 7].

Embedded RAM blocks (BRAMs) are widely applied in FPGAs to satisfy the needs of storing data. Compared with the traditional memory, BRAMs are smaller and more distributed. Furthermore, the depth and width of BRAMs can be configured with different ratios, which makes them more flexible to apply in the design of the digital system [8]. Nonetheless, with the continuous decrease in the feature size and increase

[*] Supported by the National Natural Science Foundation of China (No. 11079045, 11179003 and 11305233)

1)  E-mail:suhong@impcas.ac.cn

in the memory density, the reliability of BRAMs is on the hazard due to radiation-induced SEEs [9]. Hence, it is quite necessary for the designers to dispose some mitigation techniques for meeting the reliability constraints.

Error correcting codes (ECCs), as a mitigation technique, have been frequently adopted in the memory design to improve reliability and tolerate faults [10-12]. In this paper, the following codes will be studied: the Hamming codes and the Bose, Chaudhuri, and Hocquenghem (BCH) codes. The two codes have been studied in some literature. Liu et.al [13] investigated Hamming codes and proposed using the codes to protect finite impulse response (FIR) filters. Naseer et.al [14, 15] and Ma et.al [16] proposed using the BCH codes to improve memory reliability and implemented the codes utilizing 90nm ASIC technology. However, few irradiation experiments on the two codes are reported to verify the mitigation effect of the codes. In this paper, the shortened Hamming codes and shortened BCH codes are investigated and implemented in flash-based FPGAs. In addition, the heavy-ion experiments are carried out at the Heavy Ion Research Facility in Lanzhou (HIRFL), the experimental results indicate that the error cross-section using the shortened Hamming codes can be reduced by two orders of magnitude compared with the device without mitigation, and no errors are discovered in the experiments for the device using the shortened BCH codes. Therefore, the rationality and effectiveness of the codes are verified by the experimental results.

The paper is organized as follows. In section II, we describe the realization methods of the codes and afterwards the structure implemented in FPGAs are introduced. In section III, the synthesis report and simulation results are presented, the heavy-ion experiments are performed and the experimental results are presented and analyzed. Finally, section IV conclusions this paper.

## 2  The codes design

ECC codes employed in this work are the shortened Hamming codes and the shortened BCH codes. The block diagram of the ECC design is shown in Fig. 1. During the writing process, the parity bits are calculated through the encoder module and appended to the input data to form a codeword. Afterwards, the codeword is written into BRAMs. During the reading process, the codeword read from BRAMs enters the decoder, which detects and corrects the errors, and gets the corrected output data. In some cases, the correctable and error flag are also generated at the same time by the decoder module. In order to make comparison and statistical analysis, the codes are designed with the same length. However, the fundamental theories of different ECCs are variant, so the design of the two codes will be discussed in more detail in the next sub-sections.

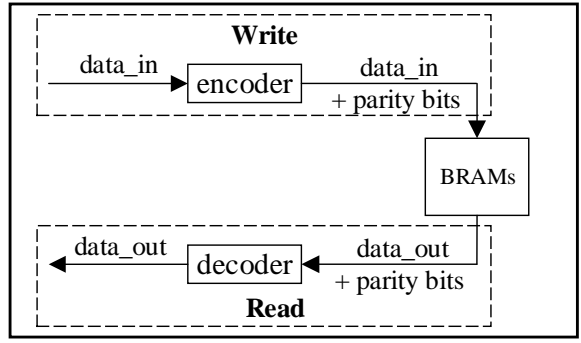

Fig.1. Block diagram of the ECC design

### 2.1 The shortened Hamming code

By properly shortening the Hamming codes discovered by Richard W. Hamming in 1950 [17], the shortened Hamming codes (known as Hsiao codes [18]) are generated and have the capacity of single error correction and double error detection (SECDED).

The (26, 20) shortened Hamming codes, which consist of 20 information bits and 6 parity bits with a minimum Hamming distance of four, are implemented in this paper. Now we assume that the codeword $A=[D_{19}D_{18}D_{17}\cdots D_1D_0P_5P_4P_3\cdots P_0]$, that is, the first 20 numbers are used to express information bits, the other numbers are used to express parity bits. If the relationship between the syndromes and the error location can be defined by Table 1, we can infer that the syndrome $S_0$ equals 1 only when the errors occur in $D_9, D_8, D_7, D_6, D_5, D_4, D_3, D_2, D_1, D_0$ and $P_0$; otherwise, the syndrome $S_0$ equals 0. Similarly, the syndromes $S_5, S_4, S_3, S_2$ and $S_1$ can also be deduced from Table 1.

Table 1. The relationship between the syndromes and the error location

| $S_5S_4S_3S_2S_1S_0$ | Error location | $S_5S_4S_3S_2S_1S_0$ | Error location |
|---|---|---|---|
| 000001 | $P_0$ | 101001 | $D_8$ |
| 000010 | $P_1$ | 110001 | $D_9$ |
| 000100 | $P_2$ | 001110 | $D_{10}$ |
| 001000 | $P_3$ | 010110 | $D_{11}$ |
| 010000 | $P_4$ | 100110 | $D_{12}$ |
| 100000 | $P_5$ | 011010 | $D_{13}$ |
| 000111 | $D_0$ | 101010 | $D_{14}$ |
| 001011 | $D_1$ | 110010 | $D_{15}$ |
| 010011 | $D_2$ | 011100 | $D_{16}$ |
| 100011 | $D_3$ | 101100 | $D_{17}$ |
| 001101 | $D_4$ | 110100 | $D_{18}$ |
| 010101 | $D_5$ | 111000 | $D_{19}$ |
| 100101 | $D_6$ | … | … |
| 011001 | $D_7$ | 000000 | No error |

Base on the above theory, the syndromes can be generated and expressed with the Eq. (1). Furthermore, the Eq. (1) can be expressed as the matrix Eq. (2). The matrix equation can be simplified as $S^T=H \cdot A^T$, where $S$ is the syndrome matrix, $H$ is the parity-check matrix, $A$ is the codeword and T is the matrix transposition.

$$\begin{cases} S_5 = D_{19} \oplus D_{18} \oplus D_{17} \oplus D_{15} \oplus D_{14} \oplus D_{12} \oplus D_9 \oplus D_8 \oplus D_6 \oplus D_3 \oplus P_5 \\ S_4 = D_{19} \oplus D_{18} \oplus D_{16} \oplus D_{15} \oplus D_{13} \oplus D_{11} \oplus D_9 \oplus D_7 \oplus D_5 \oplus D_2 \oplus P_4 \\ S_3 = D_{19} \oplus D_{17} \oplus D_{16} \oplus D_{14} \oplus D_{13} \oplus D_{10} \oplus D_8 \oplus D_7 \oplus D_4 \oplus D_1 \oplus P_3 \\ S_2 = D_{18} \oplus D_{17} \oplus D_{16} \oplus D_{12} \oplus D_{11} \oplus D_{10} \oplus D_6 \oplus D_5 \oplus D_4 \oplus D_0 \oplus P_2 \\ S_1 = D_{15} \oplus D_{14} \oplus D_{13} \oplus D_{12} \oplus D_{11} \oplus D_{10} \oplus D_3 \oplus D_2 \oplus D_1 \oplus D_0 \oplus P_1 \\ S_0 = D_9 \oplus D_8 \oplus D_7 \oplus D_6 \oplus D_5 \oplus D_4 \oplus D_3 \oplus D_2 \oplus D_1 \oplus D_0 \oplus P_0 \end{cases} \quad (1)$$

$$\begin{bmatrix} S_5 \\ S_4 \\ S_3 \\ S_2 \\ S_1 \\ S_0 \end{bmatrix} = \begin{bmatrix} 1 & 1 & 1 & 0 & 1 & 1 & 0 & 1 & 0 & 0 & 1 & 1 & 0 & 1 & 0 & 0 & 1 & 0 & 0 & 0 & 1 & 0 & 0 & 0 & 0 & 0 \\ 1 & 1 & 0 & 1 & 1 & 0 & 1 & 0 & 1 & 0 & 1 & 0 & 1 & 0 & 1 & 0 & 0 & 1 & 0 & 0 & 0 & 1 & 0 & 0 & 0 & 0 \\ 1 & 0 & 1 & 1 & 0 & 1 & 1 & 0 & 0 & 1 & 0 & 1 & 1 & 0 & 0 & 1 & 0 & 0 & 1 & 0 & 0 & 0 & 1 & 0 & 0 & 0 \\ 0 & 1 & 1 & 1 & 0 & 0 & 0 & 1 & 1 & 1 & 0 & 0 & 0 & 1 & 1 & 1 & 0 & 0 & 0 & 1 & 0 & 0 & 0 & 1 & 0 & 0 \\ 0 & 0 & 0 & 0 & 1 & 1 & 1 & 1 & 1 & 1 & 0 & 0 & 0 & 0 & 0 & 0 & 1 & 1 & 1 & 1 & 0 & 0 & 0 & 0 & 1 & 0 \\ 0 & 0 & 0 & 0 & 0 & 0 & 0 & 0 & 0 & 0 & 1 & 1 & 1 & 1 & 1 & 1 & 1 & 1 & 1 & 1 & 0 & 0 & 0 & 0 & 0 & 1 \end{bmatrix} \begin{bmatrix} D_{19} \\ D_{18} \\ D_{17} \\ \vdots \\ P_1 \\ P_0 \end{bmatrix} \quad (2)$$

Once the Eq. (1) or Eq. (2) is generated, the encoder and decoder of the (26, 20) shortened Hamming codes can be deduced easily. During the process of encoding, the syndrome matrix $S$ equals 0, which means no errors. In other words, the matrix equation can be changed as $H \cdot A^T = 0$, which can be expressed as the matrix Eq. (3). Hence, the parity-check bits $P_5$, $P_4$, $P_3$ …$P_0$ can be calculated via the matrix Eq. (3) and the codewords are generated when the parity-check bits are appended to the information bits $D_{19}$, $D_{18}$, $D_{17}$ …$D_1$, $D_0$. During the decoding process, the syndrome matrix $S$ is generated firstly via the Eq. (1) or Eq. (2). When single error happen, the error can be corrected because the number of ones in the syndrome matrix $S$ is odd and the error location can be decided according to Table 1. When double error happen, the error can be detected because the number of ones in the syndrome matrix $S$ is even.

$$\begin{bmatrix} P_5 \\ P_4 \\ P_3 \\ P_2 \\ P_1 \\ P_0 \end{bmatrix} = \begin{bmatrix} 1 & 1 & 1 & 0 & 1 & 1 & 0 & 1 & 0 & 0 & 1 & 1 & 0 & 1 & 0 & 0 & 1 & 0 & 0 & 0 \\ 1 & 1 & 0 & 1 & 1 & 0 & 1 & 0 & 1 & 0 & 1 & 0 & 1 & 0 & 1 & 0 & 0 & 1 & 0 & 0 \\ 1 & 0 & 1 & 1 & 0 & 1 & 1 & 0 & 0 & 1 & 0 & 1 & 1 & 0 & 0 & 1 & 0 & 0 & 1 & 0 \\ 0 & 1 & 1 & 1 & 0 & 0 & 0 & 1 & 1 & 1 & 0 & 0 & 0 & 1 & 1 & 1 & 0 & 0 & 0 & 1 \\ 0 & 0 & 0 & 0 & 1 & 1 & 1 & 1 & 1 & 1 & 0 & 0 & 0 & 0 & 0 & 0 & 1 & 1 & 1 & 1 \\ 0 & 0 & 0 & 0 & 0 & 0 & 0 & 0 & 0 & 0 & 1 & 1 & 1 & 1 & 1 & 1 & 1 & 1 & 1 & 1 \end{bmatrix} \begin{bmatrix} D_{19} \\ D_{18} \\ D_{17} \\ \vdots \\ D_1 \\ D_0 \end{bmatrix} \quad (3)$$

**2.2 The shortened BCH code**

The BCH codes form a large class of powerful random error-correcting cyclic codes. It can be designed with specific rules and correct more random errors [19]. The ECC used in this study is the (26, 16) shortened BCH codes. It consists of 16 information bits with a minimum distance of 5 and has error correcting capability of two errors.

The (26, 16) shortened BCH codes is constructed based on the (31, 21) BCH codes. The generator polynomial g(X) of the (31, 21) BCH codes is specified in terms of its roots from the Galois field GF ($2^5$) and can be calculated by Eq. (4)

$$g(X)=LCM\{\phi_1(X),\phi_3(X)\}=(X^5+X^2+1)(X^5+X^4+X^3+X^2+1)=X^{10}+X^9+X^8+X^6+X^5+X^3+1 \quad (4)$$

where $\varphi_i(X)$ is the minimal polynomial, LCM is the least common multiple (LCM). On the basis of g(X), the systematic generator matrix **G** can be generated with the following Eq. (5). Moreover, the Eq. (5) can be expressed as the matrix Eq. (6).

$$\mathbf{G} = \begin{bmatrix} x^{30}+x^{30} \bmod g(x) \\ x^{29}+x^{29} \bmod g(x) \\ x^{28}+x^{28} \bmod g(x) \\ \vdots \\ \vdots \\ \vdots \\ x^{11}+x^{11} \bmod g(x) \\ x^{10}+x^{10} \bmod g(x) \end{bmatrix} = \begin{bmatrix} x^{30}+x^9+x^8+x^7+x^5+x^4+x^2 \\ x^{29}+x^8+x^7+x^6+x^4+x^3+x \\ x^{28}+x^7+x^6+x^5+x^3+x^2+1 \\ \vdots \\ \vdots \\ \vdots \\ x^{11}+x^8+x^7+x^5+x^4+x^3+x+1 \\ x^{10}+x^9+x^8+x^6+x^5+x^3+1 \end{bmatrix} \quad (5)$$

$$\mathbf{G} = \begin{bmatrix}
1 & 0 & 0 & 0 & 0 & 0 & 0 & 0 & 0 & 0 & 0 & 0 & 0 & 0 & 0 & 0 & 0 & 0 & 0 & 0 & 1 & 1 & 1 & 0 & 1 & 1 & 0 & 1 & 0 & 0 \\
0 & 1 & 0 & 0 & 0 & 0 & 0 & 0 & 0 & 0 & 0 & 0 & 0 & 0 & 0 & 0 & 0 & 0 & 0 & 0 & 0 & 1 & 1 & 1 & 0 & 1 & 1 & 0 & 1 & 0 \\
0 & 0 & 1 & 0 & 0 & 0 & 0 & 0 & 0 & 0 & 0 & 0 & 0 & 0 & 0 & 0 & 0 & 0 & 0 & 0 & 0 & 0 & 1 & 1 & 1 & 0 & 1 & 1 & 0 & 1 \\
0 & 0 & 0 & 1 & 0 & 0 & 0 & 0 & 0 & 0 & 0 & 0 & 0 & 0 & 0 & 0 & 0 & 0 & 0 & 0 & 1 & 1 & 1 & 1 & 0 & 0 & 0 & 0 & 1 & 0 \\
0 & 0 & 0 & 0 & 1 & 0 & 0 & 0 & 0 & 0 & 0 & 0 & 0 & 0 & 0 & 0 & 0 & 0 & 0 & 0 & 1 & 1 & 1 & 1 & 0 & 0 & 0 & 0 & 0 & 1 \\
0 & 0 & 0 & 0 & 0 & 1 & 0 & 0 & 0 & 0 & 0 & 0 & 0 & 0 & 0 & 0 & 0 & 0 & 0 & 0 & 1 & 1 & 0 & 1 & 0 & 0 & 0 & 1 & 0 & 0 \\
0 & 0 & 0 & 0 & 0 & 0 & 1 & 0 & 0 & 0 & 0 & 0 & 0 & 0 & 0 & 0 & 0 & 0 & 0 & 0 & 0 & 1 & 1 & 0 & 1 & 0 & 0 & 0 & 1 & 0 \\
0 & 0 & 0 & 0 & 0 & 0 & 0 & 1 & 0 & 0 & 0 & 0 & 0 & 0 & 0 & 0 & 0 & 0 & 0 & 0 & 0 & 0 & 1 & 1 & 0 & 1 & 0 & 0 & 0 & 1 \\
0 & 0 & 0 & 0 & 0 & 0 & 0 & 0 & 1 & 0 & 0 & 0 & 0 & 0 & 0 & 0 & 0 & 0 & 0 & 0 & 1 & 1 & 1 & 0 & 1 & 1 & 1 & 0 & 0 \\
0 & 0 & 0 & 0 & 0 & 0 & 0 & 0 & 0 & 1 & 0 & 0 & 0 & 0 & 0 & 0 & 0 & 0 & 0 & 0 & 1 & 1 & 1 & 1 & 0 & 1 & 1 & 1 & 0 \\
0 & 0 & 0 & 0 & 0 & 0 & 0 & 0 & 0 & 0 & 1 & 0 & 0 & 0 & 0 & 0 & 0 & 0 & 0 & 0 & 1 & 1 & 1 & 1 & 0 & 1 & 1 & 1 \\
0 & 0 & 0 & 0 & 0 & 0 & 0 & 0 & 0 & 0 & 0 & 1 & 0 & 0 & 0 & 0 & 0 & 0 & 0 & 1 & 1 & 1 & 1 & 0 & 0 & 1 & 1 & 1 \\
0 & 0 & 0 & 0 & 0 & 0 & 0 & 0 & 0 & 0 & 0 & 0 & 1 & 0 & 0 & 0 & 0 & 0 & 0 & 1 & 0 & 0 & 1 & 0 & 1 & 0 & 0 & 1 & 1 \\
0 & 0 & 0 & 0 & 0 & 0 & 0 & 0 & 0 & 0 & 0 & 0 & 0 & 1 & 0 & 0 & 0 & 0 & 0 & 1 & 0 & 1 & 0 & 0 & 1 & 1 & 1 & 0 & 1 \\
0 & 0 & 0 & 0 & 0 & 0 & 0 & 0 & 0 & 0 & 0 & 0 & 0 & 0 & 1 & 0 & 0 & 0 & 0 & 0 & 1 & 0 & 1 & 1 & 1 & 1 & 1 & 0 & 1 & 0 \\
0 & 0 & 0 & 0 & 0 & 0 & 0 & 0 & 0 & 0 & 0 & 0 & 0 & 0 & 0 & 1 & 0 & 0 & 0 & 0 & 0 & 1 & 0 & 1 & 1 & 1 & 1 & 1 & 0 & 1 \\
0 & 0 & 0 & 0 & 0 & 0 & 0 & 0 & 0 & 0 & 0 & 0 & 0 & 0 & 0 & 0 & 1 & 0 & 0 & 0 & 0 & 1 & 1 & 0 & 0 & 0 & 0 & 1 & 0 & 1 & 0 \\
0 & 0 & 0 & 0 & 0 & 0 & 0 & 0 & 0 & 0 & 0 & 0 & 0 & 0 & 0 & 0 & 0 & 1 & 0 & 0 & 0 & 0 & 1 & 1 & 0 & 0 & 0 & 0 & 1 & 0 & 1 \\
0 & 0 & 0 & 0 & 0 & 0 & 0 & 0 & 0 & 0 & 0 & 0 & 0 & 0 & 0 & 0 & 0 & 0 & 1 & 0 & 0 & 1 & 1 & 0 & 1 & 1 & 1 & 0 & 1 & 1 & 0 \\
0 & 0 & 0 & 0 & 0 & 0 & 0 & 0 & 0 & 0 & 0 & 0 & 0 & 0 & 0 & 0 & 0 & 0 & 0 & 1 & 0 & 0 & 1 & 1 & 0 & 1 & 1 & 1 & 0 & 1 & 1 \\
0 & 0 & 0 & 0 & 0 & 0 & 0 & 0 & 0 & 0 & 0 & 0 & 0 & 0 & 0 & 0 & 0 & 0 & 0 & 0 & 1 & 1 & 1 & 0 & 1 & 1 & 0 & 1 & 0 & 0 & 1
\end{bmatrix} \quad (6)$$

In order to get the (26, 16) shortened BCH codes, the matrix Eq. (6) have to be processed. By deleting the first 5 rows and the first 5 columns of the matrix Eq. (6), a reduced generator matrix **G** is obtained, as shown in the matrix Eq. (7).

$$\mathbf{G} = \begin{bmatrix}
1 & 0 & 0 & 0 & 0 & 0 & 0 & 0 & 0 & 0 & 0 & 0 & 0 & 0 & 0 & 1 & 1 & 0 & 1 & 0 & 0 & 0 & 1 & 0 & 0 \\
0 & 1 & 0 & 0 & 0 & 0 & 0 & 0 & 0 & 0 & 0 & 0 & 0 & 0 & 0 & 0 & 1 & 1 & 0 & 1 & 0 & 0 & 0 & 1 & 0 \\
0 & 0 & 1 & 0 & 0 & 0 & 0 & 0 & 0 & 0 & 0 & 0 & 0 & 0 & 0 & 0 & 0 & 1 & 1 & 0 & 1 & 0 & 0 & 0 & 1 \\
0 & 0 & 0 & 1 & 0 & 0 & 0 & 0 & 0 & 0 & 0 & 0 & 0 & 0 & 1 & 1 & 1 & 1 & 0 & 1 & 1 & 1 & 0 & 0 \\
0 & 0 & 0 & 0 & 1 & 0 & 0 & 0 & 0 & 0 & 0 & 0 & 0 & 0 & 0 & 1 & 1 & 1 & 1 & 0 & 1 & 1 & 1 & 0 \\
0 & 0 & 0 & 0 & 0 & 1 & 0 & 0 & 0 & 0 & 0 & 0 & 0 & 0 & 0 & 0 & 1 & 1 & 1 & 1 & 0 & 1 & 1 & 1 \\
0 & 0 & 0 & 0 & 0 & 0 & 1 & 0 & 0 & 0 & 0 & 0 & 0 & 0 & 1 & 1 & 1 & 1 & 0 & 0 & 1 & 1 & 1 & 1 \\
0 & 0 & 0 & 0 & 0 & 0 & 0 & 1 & 0 & 0 & 0 & 0 & 0 & 0 & 1 & 0 & 0 & 1 & 0 & 1 & 0 & 0 & 1 & 1 \\
0 & 0 & 0 & 0 & 0 & 0 & 0 & 0 & 1 & 0 & 0 & 0 & 0 & 0 & 1 & 0 & 1 & 0 & 0 & 1 & 1 & 1 & 0 & 1 \\
0 & 0 & 0 & 0 & 0 & 0 & 0 & 0 & 0 & 1 & 0 & 0 & 0 & 0 & 1 & 0 & 1 & 1 & 1 & 1 & 1 & 0 & 1 & 0 \\
0 & 0 & 0 & 0 & 0 & 0 & 0 & 0 & 0 & 0 & 1 & 0 & 0 & 0 & 0 & 1 & 0 & 1 & 1 & 1 & 1 & 1 & 0 & 1 \\
0 & 0 & 0 & 0 & 0 & 0 & 0 & 0 & 0 & 0 & 0 & 1 & 0 & 0 & 0 & 1 & 1 & 0 & 0 & 0 & 0 & 1 & 0 & 1 & 0 \\
0 & 0 & 0 & 0 & 0 & 0 & 0 & 0 & 0 & 0 & 0 & 0 & 1 & 0 & 0 & 0 & 1 & 1 & 0 & 0 & 0 & 0 & 1 & 0 & 1 \\
0 & 0 & 0 & 0 & 0 & 0 & 0 & 0 & 0 & 0 & 0 & 0 & 0 & 1 & 0 & 0 & 1 & 1 & 0 & 1 & 1 & 1 & 0 & 1 & 1 & 0 \\
0 & 0 & 0 & 0 & 0 & 0 & 0 & 0 & 0 & 0 & 0 & 0 & 0 & 0 & 1 & 0 & 0 & 1 & 1 & 0 & 1 & 1 & 1 & 0 & 1 & 1 \\
0 & 0 & 0 & 0 & 0 & 0 & 0 & 0 & 0 & 0 & 0 & 0 & 0 & 0 & 0 & 1 & 1 & 1 & 0 & 1 & 1 & 0 & 1 & 0 & 0 & 1
\end{bmatrix} \quad (7)$$

Similarly, once the generator matrix **G**, or alternatively the parity-check matrix **H** is generated, the encoder and decoder of the (26, 16) shortened BCH codes can be deduced easily. Here, we assume that the codeword $B=[D_{15}D_{14}D_{13}…D_1D_0P_9P_8…P_0]$. During the encoding process, the parity-check bits can be calculated by the equation $B=[D_{15}D_{14}D_{13}…D_1D_0]·G$. For decoding purposes, conventional employed decoding schemes, such as Peterson, Berlekamp and Chain Decoding algorithms, all require a multi-cycle latency [14, 19]. To solve this problem, a similar decoding method as the above Hamming codes is adopted in this paper. The syndrome matrix **S** is generated via the equation $S^T=H·B^T$, where **H** is the parity-check matrix and can be deduced using the generator matrix **G**. In detail, the parity-check matrix **H** and the generator matrix **G** can be expressed as the forms of $H=[PI_r]$ and $G=[I_kQ]$, where **P** is the transposition of **Q**. However, the error locations corresponding to the syndrome need to be pre-computed and it is relatively complex compared with the shortened Hamming codes.

### 2.3 The structure implemented in FPGAs

Based on the above derivation, it is discovered that the encoder and decoder structure have the comparability between the shortened Hamming codes and the shortened BCH codes. As a consequence, taking BCH codes as an example, this paper shows the structure implemented in FPGAs, as shown in Fig. 2.

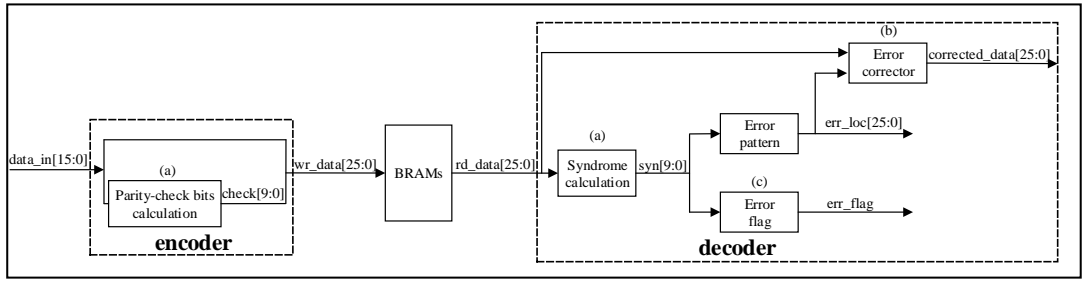

Fig.2. Block diagram of BCH codes with sub-modules: (a) Parity-check bits calculation or Syndrome calculation, (b) Error corrector, (c) Error flag.

For encoder, it is implemented through XOR operation. As shown in Fig. 3(a), the parity-check bits are calculated and appended to the data bits. The codeword is generated and stored in BRAMs. For decoder, it can be divided into the following sub-modules: syndrome calculation (Fig. 3(a)), error pattern, error corrector (Fig. 3(b)) and error flag (Fig. 3(c)). The decoding starts by reading the codeword from BRAMs. Afterwards, the syndromes are calculated through the syndrome calculation module. In fact, the circuit for the syndrome calculation, implemented through XOR operation, is similar with the parity-check bits calculation. And the syndrome should be evaluated to zero if there are no single event upsets (SEUs). But when SEUs happen, the syndrome should be nonzero. Meanwhile, the error flag can be generated through OR operation of the syndromes, as shown in Fig. 3(c). In addition, different errors will correspond to different error pattern. The error location can be determined by the error pattern which need to be computed in advance. At last, the corrected data can be generated through XOR operation between the read codeword and the error location as shown in Fig. 3(b).

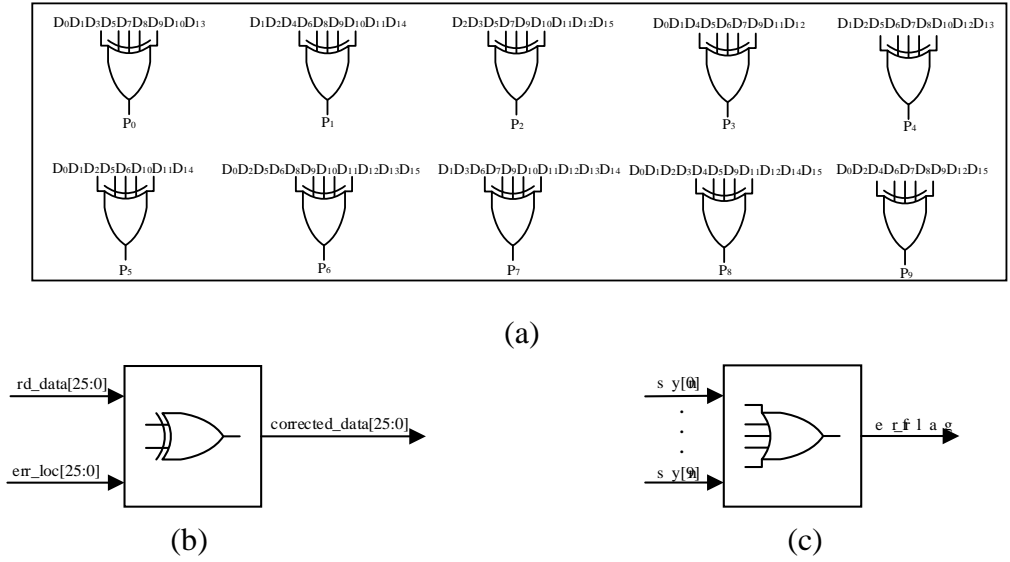

Fig.3. More detailed schematic of the sub-modules: (a) Parity-check bits calculation or Syndrome calculation, (b) Error corrector, (c) Error flag.

## 3 Results and discussion

In the previous sections, the coding methods and the structure implemented in FPGAs have been presented. In this section, the results obtained for the two codes will be discussed and analyzed. First, the FPGA synthesis report will be discussed and afterwards the simulation results will also be presented. Then, the heavy-ion experiments are performed and the experimental results will be presented and discussed. In addition, it is necessary to illustrate that the concrete design is implemented on the device APA600PQ208I, a flash-based FPGA from Microsemi's ProASIC PLUS product family. The product family is fabricated with a 0.22um feature size and 4 layer metal CMOS process. The product family has up to 1 million system gates, 88 embedded block RAMs (each consisting of 2304bits) [20].

**3.1 Synthesis and simulation**

3.1.1 Synthesis results

The design is synthesized using the Synplify Pro tool. The synthesis report of the resource occupation of the encoders and decoders have been documented in Table 2. For the encoders, the core cell usage for the shortened BCH codes is slightly higher than the shortened Hamming codes, the ratio is 1.36. But for the decoders, the core cell usage for the shortened BCH codes is much higher than the shortened Hamming codes, the ratio is 7.11. This can be explained by the fact that the different resource usage of the error pattern is the main reason. Specifically, the error pattern of the Hamming codes needs to check 26 ($C_{26}^1$) values, while the error pattern of the BCH codes needs to check 351 ($C_{26}^1 + C_{26}^2$) values. Furthermore, the calculation of the parity bits and the syndromes for the shortened BCH codes is relatively complex compared with the shortened Hamming codes. As a result, more resource is occupied for the shortened BCH codes.

Table 2. FPGA core cell usage of the encoders and decoders

| Codes | Encoder | Decoder |
|---|---|---|
| Hamming | 45 | 128 |
| BCH | 61 | 910 |

3.1.2 Simulation results

In order to verify the validity of the codes, simulations using the ModelSim tool from Mentor Graphics are carried out before the irradiation experiments. In the simulation process, the method of fault injection is adopted to simulate the upset.

As an example, a logical "checkerboard" pattern of alternating 0 and 1 value is written into the BRAMs, meanwhile, the errors including 1 bit error and 2 bits error are also injected into the BRAMs. Simulation diagrams selected from the simulation results have been illustrated in Fig. 4 and Fig. 5. For the shortened Hamming codes, 1 bit error (Fig. 4 (a)) is corrected, also, the error flag and the corrected flag are set high simultaneously to indicate that it is a correctable error. 2 bits error (Fig. 4 (b)) is detected, besides, only the error flag is set high to indicate that it is an uncorrectable error. For the shortened BCH codes, 1 bit error (Fig. 5 (a)) and 2 bits error (Fig. 5 (b)) are all corrected, and the error flag is also set high to indicate that an error has occurred. As a consequence, the effectiveness of the codes can be proven by simulations.

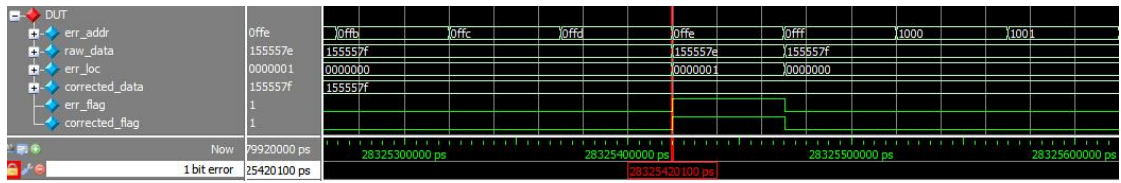

(a)

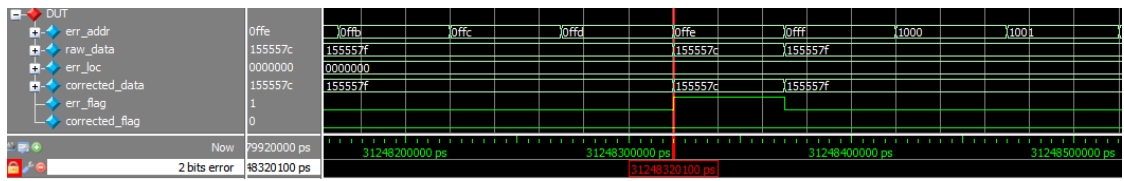

(b)

Fig.4. Simulation diagrams of the shortened Hamming codes: (a) 1 bit error, (b) 2 bits error.

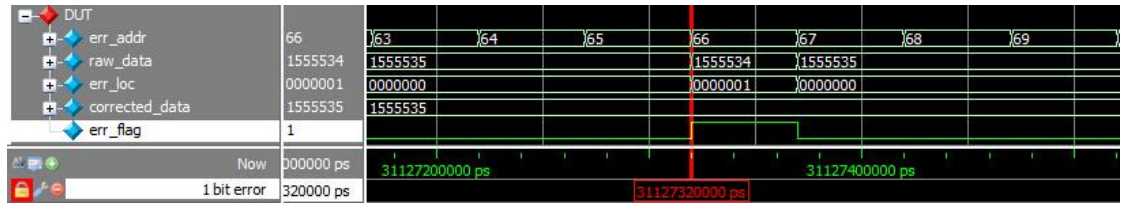

(a)

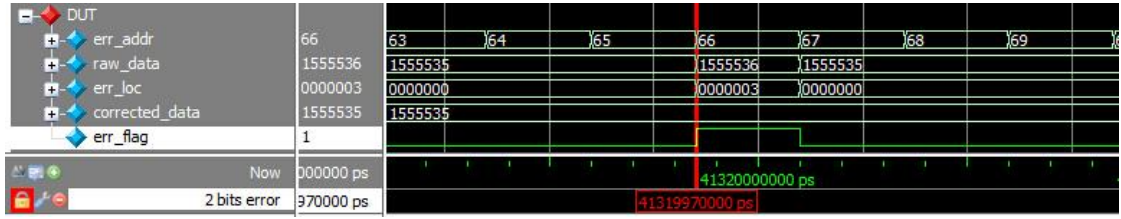

(b)

Fig.5. Simulation diagrams of the shortened BCH codes: (a) 1 bit error, (b) 2 bits error.

## 3.2 Heavy-ion experiment

3.2.1 Devices under test

In the experiment, the devices were de-capped using an acid etching machine so that heavy ions could reach the sensitive regions before the irradiation experiment, as shown in Fig. 6. In order to enhance the experimental accuracy, three pieces of test chips were selected as the experimental samples as follows: one chip, defined as DUT1 and served as a control group, was designed without mitigation; other two test chips were the main objects of observation and mitigated in terms of the above ECC designs respectively, one adopting the shortened Hamming codes is defined as DUT2 and the other adopting the shortened BCH codes is defined as DUT3.

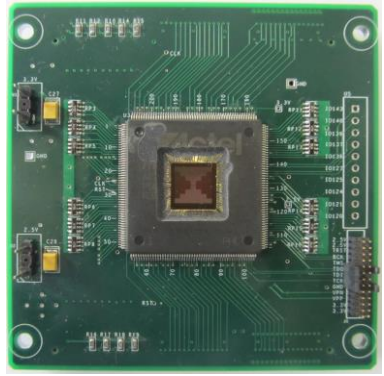

Fig.6. The picture of device under test on the daughter board

3.2.2 Description of the experimental setup

The irradiation experiments were carried out at the HIRFL, using $^{209}$Bi ions with initial energy of 9.5MeV/nucleon. The tests were performed in air and at normal incidence. A fixed Linear Energy Transfer (LET) value, 99.8 MeV cm$^2$/mg, was chosen in the experiments.

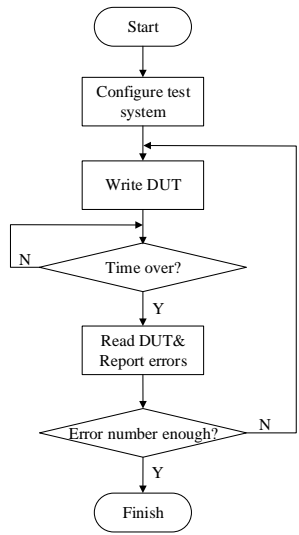

Fig.7. The flow chart in the irradiation experiments

The details of the test process are illustrated in Fig. 7. At first, the SEE test system needs to be configured. And after the system is configured successfully, the DUT is initialized a known state. In the experiments, a logical "checkerboard" pattern of alternating 0 and 1 value is adopted. The operating frequency of the DUT is set at 10MHz all the time. The time interval between the write operation and read operation is set as 120 seconds during the tests. And the following information of SEU events is recorded in real time during the experiments: error address, error time, and error data.

3.2.3 Results and analysis

The SEU cross-sections for different DUTs are illustrated in Fig. 8. We can see that the device without mitigation (DUT1) is more sensitive than the devices with mitigation (DUT2 and DUT3).

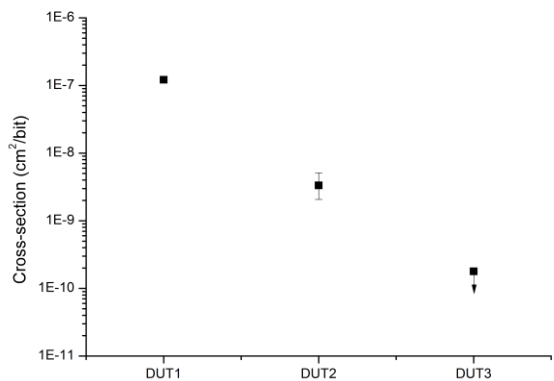

Fig.8. Cross sections for different DUTs

(Two sigma error bars are plotted for DUT1 and DUT2, but the error bar is too short to be seen for DUT1. For DUT3, the downward arrow indicates that no errors were observed at the tested fluence of 46681 ions/cm$^2$)

In detail, the bits upset event distribution for DUT1 is shown in Fig. 9, 1 bit error and 2 bits error were discovered, the corresponding cross-section is $1.17 \times 10^{-7}$ cm$^2$/bit and $4.56 \times 10^{-9}$ cm$^2$/bit, 3 bits error or more didn't appear. It reveals that 1 bit error and 2 bits error dominate the total error distribution under the given experimental condition. Furthermore, the cross-section of 2 bits error is two orders of magnitude lower than the cross-section of 1 bit error. For DUT2, the error cross-section is at the order of $10^{-9}$ cm$^2$/bit and reduced by two orders of magnitude compared with DUT1. For DUT3, no errors were discovered during the course of the experiment. The experiment results show that the mitigation methods are effective for the selected FPGAs. On the other hand, it can be seen that the shortened BCH codes are more effective than the shortened Hamming codes.

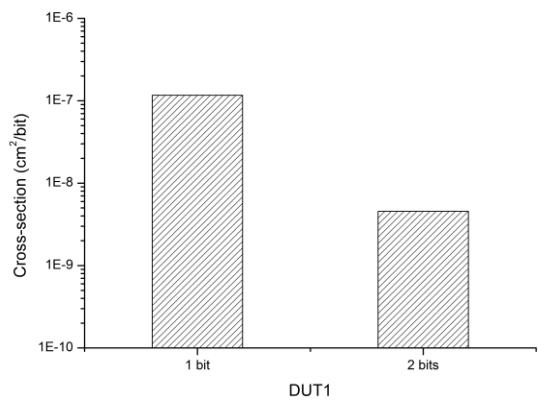

Fig.9. The bits upset event distribution for DUT1

## 4 Conclusions

In this paper, two different ECC mitigation designs for BRAMs have been investigated and implemented in flash-based FPGAs. The synthesis report and simulation results have been presented. Meanwhile, the heavy-ion experiments using $^{209}$Bi ions are carried out at the HIRFL. For the device mitigated by the shortened Hamming codes, the error cross-section is reduced by two orders of magnitude compared with the device without mitigation; for the device mitigated by shortened BCH codes, no errors are discovered in the experiments. By the above results, the rationality and effectiveness of the codes have been verified successfully. In addition, the research may provide some references for the relevant semiconductor manufacturers to improve the efficiency and reliability of products.

**Acknowledgements**

This work was supported by the National Natural Science Foundation of China (No. 11079045, 11179003 and 11305233). The SEE test in this paper was conducted at the Heavy Ion Research Facility in Lanzhou (HIRFL) located at Institute of Modern Physics, Chinese Academy of Sciences (IMPCAS). The authors are grateful for the support of the HIRFL team.

**References**

[1] Rezgui S, Wang J J, Sun Y et al. IEEE Trans. Nucl. Sci., 2008, **55**: 3328-3335
[2] Battezzati N, Gerardin S, Manuzzato A et al. IEEE Trans. Nucl. Sci., 2009, **56**: 3534-3541
[3] YANG S Y, CAO Z, DA D A et al. Chin. Phys. C, 2009, 33 (05): 369-373
[4] Dodd P E, Shaneyfelt M R, Schwank J R et al. IEEE Trans. Nucl. Sci., 2010, **57**: 1747-1763
[5] Nicolaidis M. Soft errors in modern electronic systems. Springer Science & Business Media, 2010
[6] Rezgui S, McCollum J, Won R et al. IEEE Trans. Nucl. Sci., 2010, **57**: 3716-3724
[7] Rezgui S, Wang J J, Tung E C et al. IEEE Trans. Nucl. Sci., 2007, **54**: 2512-2524
[8] Li Y B, Nelson B, Wirthlin M et al. IEEE Trans. Nucl. Sci., 2013, **60**: 2720-2727
[9] Varghese B, Sreelal S, Vinod P et al. Information & Communication Technologies (ICT), 2013 IEEE Conference on, 2013. 1086-1090
[10] Chen P Y, Yeh YT, Chen C H et al. in Test Conference, 2006. ITC '06. IEEE International, 2006. 1-10
[11] Namba K, Pontarelli S, Ottavi M et al. IEEE Trans. Devi. Mate Reliability., 2014, **14**: 664-671
[12] Liu S, Sorrenti G, Reviriego P et al. IEEE Trans. Nucl. Sci., 2012, **59**: 619-624
[13] Liu S F, Reviriego P, Maestro J A et al. IEEE Trans. Nucl. Sci., 2010, **57**: 2112-2118
[14] Naseer R, Draper J. in Electronics, Circuits and Systems, 2008. ICECS 2008. 15th IEEE International Conference on, 2008. 586-589
[15] Naseer R, Draper J. in Solid-State Circuits Conference, 2008. ESSCIRC 2008. 34th European, 2008. 222-225
[16] Ma W J, Cui X L, Lee C L. in ASIC (ASICON), 2013 IEEE 10th International Conference on, 2013, 1-4
[17] Hamming R W. Bell System technical journal., 1950, **29**: 147-160
[18] Hsiao M Y. IBM Journal of Research and Development., 1970, **14**: 395-401
[19] Lin S, Costello D J. Error Control Coding. Second edition. Prentice-Hall, 2004
[20] Microsemi Corporation. ProASIC PLUS Flash Family FPGAs Datasheet, 2010. http: www.microsemi.com